\documentclass[preprint2]{aastex}
\usepackage{here}
\shorttitle{Cosmic ray anisotropy from EAS-TOP}
\shortauthors{The EAS-TOP Collaboration}

\begin{document}

\title{Evolution of the cosmic ray anisotropy above $10^{14}$ eV}

\author{M. Aglietta\altaffilmark{1,2}, V.V. Alekseenko\altaffilmark{3}, B. Alessandro\altaffilmark{2}, P. Antonioli\altaffilmark{4},
F.~Arneodo\altaffilmark{5}, L.~Bergamasco\altaffilmark{2,6}, M.~Bertaina\altaffilmark{2,6}, R.~Bonino\altaffilmark{1,2},
A.~Castellina\altaffilmark{1,2}, A.~Chiavassa\altaffilmark{2,6},
B.~D'Ettorre~Piazzoli\altaffilmark{7}, G.~Di~Sciascio\altaffilmark{7,8},
W.~Fulgione\altaffilmark{1,2}, P.~Galeotti\altaffilmark{2,6},
P.L.~Ghia\altaffilmark{1,5,9}, M.~Iacovacci\altaffilmark{7}, G.~Mannocchi\altaffilmark{1,2},
C.~Morello\altaffilmark{1,2}, G.~Navarra\altaffilmark{2,6},
O.~Saavedra\altaffilmark{2,6}, A.~Stamerra\altaffilmark{6,10},
G.C.~Trinchero\altaffilmark{1,2}, S.~Valchierotti\altaffilmark{2,6},
P.~Vallania\altaffilmark{1,2}, S.~Vernetto\altaffilmark{1,2},
C.~Vigorito\altaffilmark{2,6} \\(The EAS-TOP Collaboration)}

%All the affiliations.
\altaffiltext{1} {Istituto di Fisica dello Spazio Interplanetario, INAF, Torino, Italy}
\altaffiltext{2} {Istituto Nazionale di Fisica Nucleare, Torino, Italy}
\altaffiltext{3} {Institute for Nuclear Research, AS Russia, Baksan Neutrino Observatory, Russia}
\altaffiltext{4} {Istituto Nazionale di Fisica Nucleare, Bologna, Italy}
\altaffiltext{5} {Laboratori Nazionali del Gran Sasso, INFN, Assergi (AQ), Italy}
\altaffiltext{6} {Dipartimento di Fisica Generale dell'Universit\`a, Torino, Italy}
\altaffiltext{7} {Dipartimento di Scienze Fisiche dell'Universit\`a and INFN, Napoli, Italy}
\altaffiltext{8} {Presently at Istituto Nazionale di Fisica Nucleare, Roma Tor Vergata, Italy}
\altaffiltext{9} {Presently at Institut de Physique Nucleaire, CNRS, Orsay, France}
\altaffiltext{10} {Presently at Dipartimento di Fisica dell'Universit\`a and INFN, Pisa, Italy}

\begin{abstract}
The amplitude and phase of the cosmic ray anisotropy are well established experimentally between $10^{11}$ eV and $10^{14}$ eV. The study of their evolution 
into the energy region $10^{14}-10^{16}$ eV can provide a significant tool for the understanding of the steepening (``knee'') of the primary spectrum.
In this letter we extend the EAS-TOP measurement performed at $E_0 \approx 10^{14}$ eV, to higher energies by using the full data set (8 years of data taking).
Results derived at about $10^{14}$ and $4\cdot10^{14}$ eV
 are compared and discussed.
Hints of increasing amplitude and change of phase above $10^{14}$ eV are reported.
The significance of the observation for the understanding of cosmic ray propagation is discussed.
\end{abstract}


\section{Introduction}

The steepening (``knee'') observed at $E_0\approx3\cdot 10^{15}$ eV represents a main feature of the energy spectrum of cosmic rays and its characterization is therefore a main tool for the understanding of the galactic radiation.
Composition studies have shown that it is related to the steepening of the lightest primaries (protons, helium, CNO) spectra 
\citep{etcomp,kcomp}. 

Such effect can be due, on the one side, to energy limits of the acceleration process at the source, namely diffusive shock acceleration in supernova remnants, generally considered to be the sources of galactic cosmic rays. The maximum energy of the accelerated protons is, indeed, calculated to occur in the $10^{15}$ eV energy region \citep{berez,berezvolk}, but could reach up to about $ 10^{17}$ eV \citep{ptuskin}.
On the other side, this feature has been possibly explained in terms of a change in the cosmic ray propagation properties inside the Galaxy \citep{pet,zat}. 
Galactic propagation is described through diffusion models whose parameters have been obtained through composition studies (mainly from the ratio of secondary to primary nuclei) at energies well below 1 TeV (see e.g. \citep{jones,strong}). The diffusion coefficient, $D$, is found to increase with magnetic rigidity ($D \propto R^{0.6}$, or $D \propto R^{0.3}$ for models including reacceleration). 
However, no confirmation, and no information has till now been obtained at higher energies, where the main observable is represented by the large scale anisotropy in the cosmic rays arrival directions, that is known to be strictly related to the diffusion coefficient (see e.g. \citep{bg}). 
The study of the evolution of the anisotropy in the ``knee'' energy region can therefore provide a significant test of the diffusion models, and a valuable insight for the discrimination between the two possible explanations of the spectral steepening.

At $E_0\approx10^{14}$ eV the EAS-TOP{\footnote {The Extensive Air Shower array on TOP of the Gran Sasso underground laboratories.} results \citep{etapj} demonstrated that the 
 main features of the anisotropy (i.e. of  cosmic ray propagation) are similar to the ones measured at lower energies ($10^{11} \div 10^{14}$ eV), both with respect to amplitude 
($(3 \div 6)\cdot 10^{-4}$) and phase ((0 $\div $ 4) h LST)
\citep{musala, poatina, norikura, baksan1, baksan2, MACRO, kamioka, tibet, superk, milagro}.
At higher energies the limited statistics does not allow to draw any firm conclusion \citep{akeno, gherardy, kascadean, tibetsci, kgrande}.

In this letter we present the EAS-TOP measurement based on the full data-set and we extend the analysis to about $4\cdot10^{14}$ eV. 

\section{The experiment and the analysis}

The EAS-TOP Extensive Air Shower array was located at Campo Imperatore (2005 m
a.s.l., lat. $42^\circ~ 27'$~N, long. $13^\circ~ 34'$~E, INFN Gran Sasso National
Laboratory). The electromagnetic detector (used for the present analysis) \citep{aglio}
consisted of 35 modules of scintillator counters, 10 m$^2$ each, distributed over an area of about 10$^5$ m$^2$. The trigger was provided by the coincidence of any four neighbouring modules (threshold $n_p\approx 0.3$ m.i.p./module), the event rate being $f\approx$ 25 Hz. The data under discussion have been collected between January 1992 and
December 1999 for a total of 1431 full days of operation.

To select different primary energies, a cut is applied to the events based on the number of triggered modules (see table 1). The average primary energies are evaluated for primary protons and QGSJET01 hadron interaction model \citep{qgsjet} in CORSIKA \citep{kna}.

\begin{table}[H]
\begin{center}
\caption{{\it Characteristics of the two classes of events used in the
    analysis: number of triggered modules, primary energy and number of
    collected events in the East+West sectors.}}
\begin{tabular}{cccc}
\tableline\tableline
Class & $N_{modules}$ & $E_0$ [eV] & $N_{EW}$\\
\tableline
I & $\ge 4$ & $1.1\cdot 10^{14}$ & $1.5\cdot 10^9$\\
II & $\ge 12$ & $3.7\cdot 10^{14}$ & $1.7\cdot 10^8$\\
\tableline\tableline
\end{tabular}
\end{center}
\end{table}

For the analysis of the 
anisotropy, we adopt a method based on the counting rate differences between East-ward and West-ward directions, that allows to remove counting rate variations of atmospheric origin.
The events used in the analysis (see table 1) are the ones with azimuth angle $\phi$ inside $ \pm 45^\circ$ around the East and West directions, and zenith angle $\theta<40^\circ$.
The difference between the number of counts measured from the East sector, $C_E(t)$, and from the West one, $C_W(t)$, at time $t$ in a fixed interval ($\Delta t=20$ min), is related to the first derivative of the intensity $I(t)$  as:  $\frac{dI}{dt}\simeq{D(t) = }\frac{C_E(t)-C_W(t)}{\delta t}$
where $\delta t$ is the average hour angle between the vertical and each of the two sectors (1.7 h in our case).
The harmonic analysis is performed on the differences {\small$D(t)$};
the amplitudes and phases of the variation of $I(t)$ are obtained through the integration of the corresponding terms of the Fourier series \citep{etmerida}.

\section{Results}
\begin{table*}[!t]
\tabletypesize{\scriptsize}
\begin{center}
\begin{tabular}{c|ccc|ccc|ccc}
\tableline\tableline\\
$E_0$[eV] & $A^I_{sol}~10^4$ & $\phi^I_{sol}$[h] & $P^I_{sol}$(\%) &
$A^I_{sid}~10^4$ & $\phi^I_{sid}$[h] & $P^I_{sid}$(\%) & $A^I_{asid}~10^4$ &
$\phi^I_{asid}$[h] & $P^I_{asid}$(\%)\\[0.3cm]
\tableline
$1.1\cdot 10^{14}$ & $2.8\pm0.8$ & $6.0\pm1.1$ & $0.2$ & $2.6\pm0.8$ & $0.4\pm1.2$ & $0.5$ &
$1.2\pm0.8$ & $23.9\pm2.8$ & $32.5$ \\
$3.7\cdot 10^{14}$ & $3.2\pm2.5$ & $6.0\pm3.4$ & $44.1$ & $6.4\pm2.5$ & $13.6\pm1.5$ & $3.8$ &
$3.4\pm2.5$ & $22.3\pm3.2$ & $39.7$ \\
\tableline\tableline\\
& $A^{II}_{sol}~10^4$ & $\phi^{II}_{sol}$[h] & $P^{II}_{sol}$(\%) &
$A^{II}_{sid}~10^4$ & $\phi^{II}_{sid}$[h] & $P^{II}_{sid}$(\%) & $A^{II}_{asid}~10^4$ &
$\phi^{II}_{asid}$[h] & $P^{II}_{asid}$(\%)\\[0.3cm]
\tableline
$1.1\cdot 10^{14}$& $1.4\pm0.8$ & $7.0\pm1.2$ & $21.6$ & $2.3\pm0.8$ & $6.3\pm0.7$ & $1.6$ &
$0.6\pm0.8$ & - & $75.5$ \\
$3.7\cdot 10^{14}$ & $1.7\pm2.5$ & - & $79.4$ & $1.5\pm2.5$ & - & $83.5$ &
$1.2\pm2.5$ & - & $89.1$ \\
\tableline\tableline
\end{tabular}
\caption{{\it Results of the analysis of the first (amplitude $A^I$, phase $\phi^I$, and Rayleigh imitation probability $P^I$) and second harmonic ($A^{II}$, $\phi^{II}$, $P^{II}$) in solar (columns 2-4), sidereal (columns 5-7), and anti-sidereal time (columns 8-10). Phases are not defined when amplitudes are smaller than their uncertainties.}}
\end{center}
\vspace{-0.5cm}
\normalsize
\end{table*}
The harmonic analysis has been performed in solar, sidereal and anti-sidereal time{\footnote {The anti-sidereal time is a fictitious time scale symmetrical to the sidereal one with respect to the solar time and that reflects seasonal influences \citep{farleystorey}.}}. We describe in sect 3.1 the results of the analysis, while in sect. 3.2 we show the related counting rate curves.

\subsection{The harmonic analysis}

For the two different primary energies, the reconstructed amplitudes and phases of the first and second harmonics are shown in table 2, together with the corresponding Rayleigh imitation probabilities ($P$).

-- Concerning the {\bf first harmonic}:

{\bf (a)} At {\bf $1.1\cdot 10^{14}$ eV}, from the analysis in {\underline {solar time}}, 
the obtained  
amplitude and phase ($A^I_{sol}=(2.8 \pm 0.8)\cdot 10^{-4}$,
$ \phi^I_{sol}=(6.0 \pm 1.1)$ h, $P^I_{sol}=0.2\%$) 
are in excellent agreement with the expected ones from the
Compton-Getting effect \citep{CG} due
to the revolution of the Earth around the Sun: at our latitude
$A_{sol,CG}$=$3.0\cdot 10^{-4}$, $\phi_{sol,CG}$=$6.0$ h.

With respect to the {\underline {sidereal time}} analysis, the measured amplitude and phase ($A^I_{sid}=(2.6 \pm 0.8)\cdot 10^{-4}$,
$ \phi^I_{sid}=(0.4 \pm 1.2)$ h LST), with  imitation probability $P^I_{sid}=0.5\%$,
confirm the previous
EAS-TOP  result \citep{etapj}. 

Bi-monthly vectors representing the first harmonic are shown in figure \ref{fig:solarrotation} (dots), together with the expected ones (stars) from the measured solar and sidereal amplitudes. 
The expected anti-clockwise rotation of the vector is
clearly visible, showing that, at any time, the composition of the two vectors is observed, and that the expected and measured individual values
are fully compatible within the statistical uncertainties. 

{\bf (b)} At {\bf $3.7\cdot 10^{14}$ eV} the amplitude and phase of the measured first harmonic in {\underline {solar time}} are still consistent with the expected ones for the solar Compton-Getting effect, although, due to the reduced statistics, the chance imitation probability is rather high.

\begin{figure}[H]
\vspace{-1cm}
\begin{center}
\noindent
\includegraphics [width=0.31\textwidth]{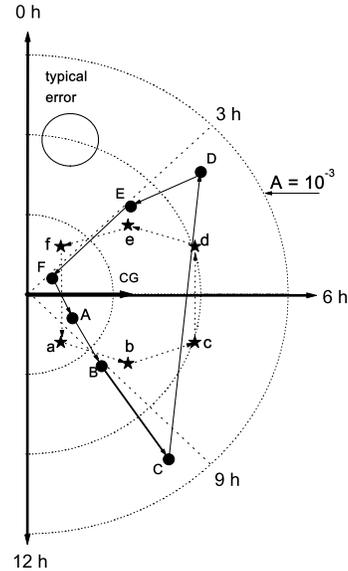}
\caption{\it Bi-monthly solar vectors at $1.1\cdot 10^{14}$ eV.
Capital and small letters refer respectively to
the expected and theoretical points (A,a = Jan + Feb; B,b = Mar + Apr; C,c = May + Jun; D,d = Jul + Aug; E,e = Sep + Oct; F,f = Nov + Dec). The typical vector statistical uncertainty is shown.
}
\label{fig:solarrotation}
\end{center}
\end{figure}
Concerning the analysis in {\underline {sidereal time}}, we obtain $A^I_{sid}=(6.4 \pm 2.5)\cdot 10^{-4}$, $ \phi^I_{sid}=(13.6 \pm 1.5)$ h LST, with an imitation probability of about 3.8\%. 
This indicates therefore a change of phase (from 0.4 to 13.6 h) and an increase of amplitude (by a factor 2.5) with respect to the first harmonic measured at $1.1\cdot 10^{14}$ eV.

-- Concerning the {\bf second harmonic} most significant ($P^{II}_{sid}=1.6\%$) is the amplitude observed in sidereal time
in the lower energy class of events (comparable with the first harmonic one:
$A^{II}_{sid}=(2.3 \pm 0.8)\cdot 10^{-4}$, $ \phi^{II}_{sid}=(6.3 \pm 0.7)$ h
LST) (see also \citep{baksan1}).

Both at $1.1\cdot 10^{14}$ eV and $3.7\cdot 10^{14}$ eV, no significant amplitude is observed in {\underline {anti-sidereal time}}, showing that no additional correction is required due to residual seasonal effects.

\subsection{The counting rate curves}
\begin{figure}[H]
\begin{center}
\vspace{-0.3cm}
\includegraphics[width=0.49\textwidth]{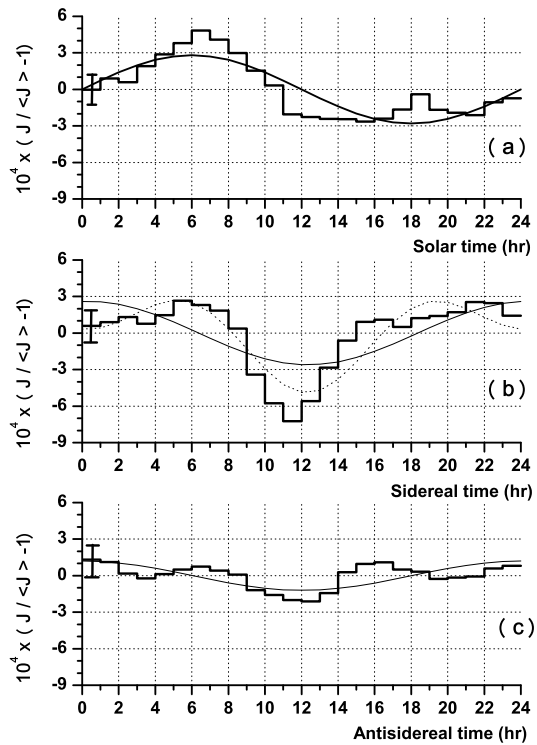}
\vspace{-0.3cm}
\caption{\it Thick black lines: counting rate curves in solar (a), sidereal (b) and anti-sidereal (c) time at $1.1\cdot 10^{14}$ TeV. The statistical uncertainty for each bin is given in the first one.
The curves resulting from the first harmonic analysis are also shown (light black lines); for the sidereal time curve, the combination of first and second harmonic (dotted black line) is additionally superimposed.}
\label{fig:ewt1}
\end{center}
\end{figure}
Besides the harmonic analysis, it is interesting to visualize the variations of the cosmic ray intensity versus time, $I(t)$, as reconstructed by integration of the East-West differences, $D(t)$. They  are shown in figs.  \ref{fig:ewt1} and  \ref{fig:ewt2}, for the classes of events at $1.1\cdot 10^{14}$ eV and $3.7\cdot 10^{14}$ eV, respectively (a,b,c for solar, sidereal, and anti-sidereal time scales).

As already shown by the harmonic analysis, at both energies the curves in solar time are dominated by the Compton-Getting effect  due to the motion of the Earth, and no modulation is visible in the anti-sidereal time scale.

A main difference is observed in the sidereal time curves: while 
the shape of the curve at $1.1\cdot 10^{14}$ eV is in remarkable agreement with the EAS and muon measurements reported at and below 100 TeV, the curve related to the highest energy class of events is characterized by a broad excess around 13-16 h LST.

\begin{figure}[H]
\begin{center}
\vspace{-0.7cm}
\includegraphics[width=0.5\textwidth]{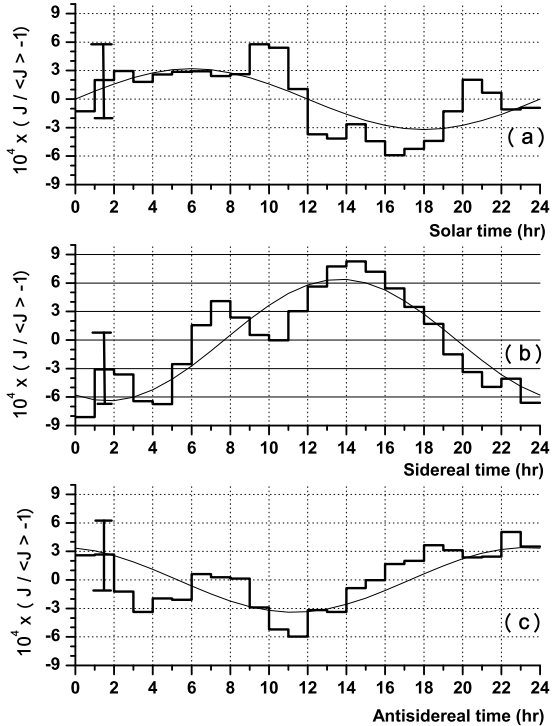}
\vspace{-0.8cm}
\caption{\it Thick black lines: counting rate curves in solar (a), sidereal (b) and anti-sidereal (c) time at $3.7\cdot 10^{14}$ TeV. The curves resulting from the first harmonic analysis are also shown (light black lines).}
\label{fig:ewt2}
\end{center}
\end{figure}

\section{Conclusions}
High stability data obtained from long time observations (8 years) from the EAS-TOP array confirm the amplitude and phase
of the cosmic ray anisotropy already reported at $10^{14}$ eV: $A^I_{sid}=(2.6 \pm 0.8)\cdot 10^{-4}$, $ \phi^I_{sid}=(0.4 \pm 1.2)$ h LST, with Rayleigh imitation probability $P^I_{sid}=0.5\%$.
The result is supported by the observation of the Compton-Getting 
effect due to the revolution of the Earth around the Sun, and by the absence of anti-sidereal effects.
It confirms the homogeneity of the anisotropy data over the energy range $10^{11}$-$10^{14}$ eV. 

At higher energies (around $4\cdot 10^{14}$ eV) the observed anisotropy shows a larger amplitude, $A^I_{sid}=(6.4 \pm 2.5)\cdot 10^{-4}$, and a different phase, $ \phi^I_{sid}=(13.6 \pm 1.5)$ h LST, with an imitation probability of 3.8\%. The statistical significance is still limited, but the measurement has the highest sensitivity with respect to previous experiments at these energies, and it is not in contradiction with any of them. 

The dependence of the anisotropy amplitude over primary energy ($A \propto E_0^\delta$) deduced from the present two measurements can be represented by a value of  $\delta = 0.74 \pm 0.41$. Therefore, at least in the energy range $(1-4) \cdot 10^{14}$ eV, such dependence is compatible with that of the diffusion coefficient as derived by composition measurements at lower energies.

On another side, the sharp increase of the anisotropy above $10^{14}$ eV may be indicative of a sharp evolution of the propagation properties, and therefore of the diffusion coefficient  just approaching the steepening of the primary spectrum.
This opens the problems of obtaining an improved theoretical and experimental description of the whole evolution of the diffusion processes vs primary energy, and understanding  how such evolution could affect the energy spectra at the "knee". From the experimental point of view, the extension of the anisotropy measurements with high sensitivity to and above $10^{15}$ eV will be of crucial significance.

\acknowledgments
V.V.A. is grateful to the INFN Gran Sasso National Laboratory for financial support through FAI funds. P.L.G. acknowledges the financial support by the European Community 7th Framework Program through the Marie Curie Grant PIEF-GA-2008-220240.

\end{document}